\documentclass[10pt,a4paper]{article}
\usepackage[latin1]{inputenc}
\usepackage{textcomp}
\usepackage{amsmath}
\usepackage{amsfonts}
\usepackage{amssymb}
\usepackage{graphicx}
\usepackage{hyperref} 
\usepackage{tikz} 
\usepackage{authblk} 
\usepackage{tocbibind} 
\newcommand{\dd}{\mathop{}\,\mathrm{d}} 
\newcommand{\de}{\mathop{}\! \Delta E} 

\author[1,2]{Bernard Derrida} 
\author[3]{Peter Mottishaw}
                    
	\affil[1]{Coll\`{e}ge de France,	11 Place Marcelin Berthelot, 75005 Paris, France}
	\affil[2]{LPS,  \'{E}cole Normale Sup\'{e}rieure, 24 rue Lhomond, 75005 Paris, France}
	\affil[3]{School of Physics and Astronomy, University of Edinburgh, James Clerk Maxwell Building, Edinburgh EH9~3FD, UK}

\title{Finite size corrections to the Parisi overlap function in the GREM}

\begin{document}
\maketitle
\begin{abstract}
	We investigate the effects of finite size corrections on the overlap probabilities in the Generalized Random Energy Model (GREM) in two situations where replica symmetry is broken in the thermodynamic limit. Our calculations do not use replicas, but shed some light on what the replica method should give for finite size corrections. In the gradual freezing situation, which is known to exhibit full replica symmetry breaking, we show that the finite size corrections lead to a modification of the simple relations between the sample averages of the overlaps $ Y_k $ between $ k $ configurations predicted by replica theory. This can be interpreted as fluctuations in the replica block size with a \emph{negative} variance. The mechanism is similar to the one we found recently in the random energy model~\cite{Derrida_2015_Finite}. We also consider a simultaneous freezing situation, which is known to exhibit one step replica symmetry breaking. We show that finite size corrections lead to full replica symmetry breaking and give a more complete derivation of the results presented in~\cite{Derrida_2016_Genealogy} for the directed polymer on a tree.
	
	\smallskip
	\noindent \textbf{Keywords.} mean field spin glasses, replica symmetry breaking, directed polymers.
\end{abstract}

	It is our pleasure to dedicate this work to our colleagues and friends J{\"u}rg Fr{\"o}hlich, Thomas Spencer and Herbert Spohn on the occasion of their 70th Birthdays.

\section{Introduction}
	Trying to predict finite size corrections of disordered systems which exhibit broken replica symmetry has been a long standing problem \cite{Young_1983_Direct,Nieuwenhuizen_1996_puzzle,Ferrero_1996_Fluctuations,Campellone_2009_Replica}. If one uses the replica approach the difficulty comes from the fact that one needs to deal with a quadratic form around a saddle point in a space with a non integer number of dimensions.  In order to shed some light on this problem, we have investigated some  exactly soluble models for which the Parisi overlap function (\cite{Mezard_1984_Nature,Mezard_1984_Replica,Mezard_1987_Spin} and references therein) can be calculated without using the	replica method. This was done in two recent works for the random energy model (REM) \cite{Derrida_2015_Finite} and for the problem of the directed polymer on a tree \cite{Derrida_2016_Genealogy}. Both models exhibit a one step broken symmetry of replica \cite{Derrida_1980_Random,Derrida_1981_Random,Gross_1984_simplest}. Despite the fact that, in the thermodynamic limit, these two models have the same free energy and the same overlap function, their finite size  corrections are quite different. In the case of the random energy model, we showed that these finite size corrections of the overlap function can be interpreted using the replica method a la Parisi, if one allows the size of the blocks to fluctuate with a {\it negative} variance \cite{Derrida_2015_Finite}, while in the directed polymer problem on a tree, finite size corrections transform one step replica symmetry breaking into full replica symmetry breaking~\cite{Derrida_2016_Genealogy}.

	In the present work we investigate the Generalized Random Energy Model (GREM) \cite{Derrida_1985_generalization} in two situations: a gradual freezing situation which is known to present a full replica symmetry breaking \cite{Derrida_1986_Magnetic,Bovier_2004_Derridas,Bovier_2004_Derridasa,Obuchi_2010_Replica} and the simultaneous freezing situation which is very similar to the directed polymer problem and for which we give a more extensive derivation of the results presented in \cite{Derrida_2016_Genealogy}.\\
	
Our main results can be expressed in rather general terms using the functions $ Y_k(q) $ introduced in the context of Parisi's  replica symmetry breaking scheme~\cite{Mezard_1984_Nature,Mezard_1984_Replica,Mezard_1987_Spin,Derrida_1997_random}. $ Y_k(q) $ is defined to be the probability that $ k $ replicas with the same quenched disorder have an overlap of at least $ q $. If $ q_{\mathcal{C},\mathcal{C'}} $ is the overlap between a pair of configurations and $ E_{\mathcal{C}} $ is the energy of a configuration, then $ Y_k(q) $ is therefore defined by
\begin{equation}
Y_k(q)=\frac{\sum_{\mathcal{C}_1} \cdots \sum_{\mathcal{C}_k} 
	\prod_{1\le i<j \le k} \Theta(q_{\mathcal{C}_i,\mathcal{C}_j}-q)\: 
		e^{-\beta (E_{\mathcal{C}_1}+ \cdots + E_{\mathcal{C}_k})}}{Z^k}
\end{equation}
where $ Z $ is the partition function ($ Z= \sum_{\mathcal{C}} e^{-\beta E_{\mathcal{C}}} $). (According to the theory of mean field spin glasses the $ Y_k(q) $ are in general sample dependent and have sample to sample fluctuations of order 1 with statistics predicted by the theory~\cite{Mezard_1984_Replica,Derrida_1997_random}. For example
\begin{equation}
\left\langle Y_2^2(q) \right\rangle = \frac{\left\langle Y_2(q) \right\rangle + 2 \, \left\langle Y_2(q) \right\rangle^2}{3}
\end{equation}
where the $ \left\langle \cdots \right\rangle  $ represent an average over quenched disorder.)
One of the outcomes of the Parisi theory is that, independent of all parameters which characterise the model  and on which the $ Y_k $ depend (temperature, magnetic field, overlap $ q $, ...), when averaged over all the samples, there are simple relations~\cite{Mezard_1984_Replica,Derrida_1997_random,Ruelle_1987_Mathematical,Bolthausen_1998_Ruelles} between the sample overlaps of the $ Y_k $  of the form (see section~\ref{sec:Recursion})
\begin{equation} \label{eq:General:Yk_Fk}
\langle Y_k \rangle = F_k(\langle Y_2 \rangle)
\end{equation}
where
\begin{equation} \label{eq:General:Fk_definition}
F_k(z) = \frac{\Gamma(k-1+z)}{\Gamma(k) \Gamma(z)}.
\end{equation}
What we will show in the present work (see section~\ref{sec:Recursion}) is that, for the gradual freezing case of the GREM, formulas \eqref{eq:General:Yk_Fk} and \eqref{eq:General:Fk_definition} are no longer valid when the leading finite size corrections are included. They instead become
\begin{equation} \label{eq:General:Yk_finite_size}
\langle Y_k \rangle = F_k(\langle Y_2 \rangle) - \epsilon \Delta_2 F_k''(\langle Y_2 \rangle) + O(\epsilon^2)
\end{equation}
where $ \epsilon \Delta_2 $ is the amplitude of the finite size correction (here $ \epsilon = O(N^{-1}) $ where $ N $ is the size of the system). This can also be interpreted as if the sample average $ \langle Y_2 \rangle $ fluctuates with a negative variance 

\begin{equation} \label{eq:General:Delta_2}
\mathrm{Var}(\left\langle Y_2 \right\rangle ) = - \epsilon \Delta_2.
\end{equation}
In terms of Parisi's replica symmetry breaking scheme (see section~\ref{sec:Recursion}) this means finite size effects can be interpreted as fluctuations in the replica block size and as in the REM case~\cite{Derrida_2015_Finite}, with a negative variance of these fluctuations.

In the simultaneous freezing case the leading finite size correction is enough to produce full replica symmetry breaking in a model which has a single step of replica symmetry breaking in the thermodynamic limit. However our results are limited to a lower order ($ \sqrt{\epsilon} = O(N^{-\frac{1}{2}})$) than the gradual freezing case and we find that the relations (\ref{eq:General:Yk_Fk},~\ref{eq:General:Fk_definition}) are satisfied at this order.

\section{Definition of the Poisson GREM}\label{sec:definition-of-the-poisson-grem}
	The generalised random energy model (GREM) was originally introduced in the context of spin glasses, but has since appeared in many guises \cite{Derrida_1985_generalization,Derrida_1986_Magnetic,Ruelle_1987_Mathematical,Bolthausen_1998_Ruelles,Bovier_2004_Derridas,Bovier_2004_Derridasa,Bolthausen_2007_Spin,Cao_2016_One}, notably to approximate polymers in a random medium \cite{Derrida_1988_Polymers,Cook_1991_Finite,Derrida_2016_Genealogy}. In this paper we will study a variant of the original GREM that we will refer to as the Poisson GREM. As in the REM \cite{Derrida_2015_Finite} this has the advantage of simplifying some of the analysis while giving the same results for large systems, including the leading finite size corrections (see appendix~\ref{sec:standard-vs-poisson-grem}).
	
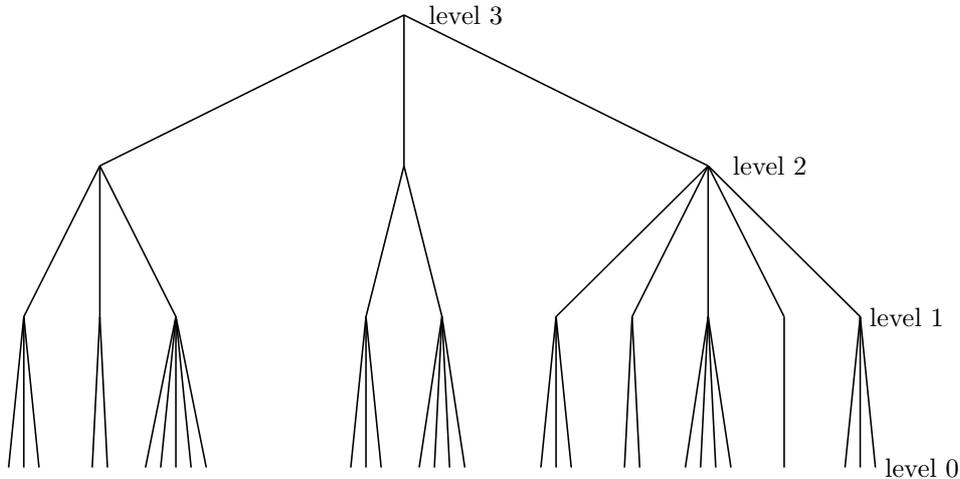
\begin{figure}
	\centering
\begin{tikzpicture}[scale=4, edge/.style={semithick}]
\draw [edge] (0,0) -- (1,-0.5); 
	\draw [edge] (1,-0.5) -- (1.5,-1); 
		\draw [edge] (1.5,-1) -- (1.55,-1.5); 
		\draw [edge] (1.5,-1) -- (1.5,-1.5);
		\draw [edge] (1.5,-1) -- (1.45,-1.5);
	\draw [edge] (1,-0.5) -- (1.25,-1);
		\draw [edge] (1.25,-1) -- (1.25,-1.5);
	\draw [edge] (1,-0.5) -- (1,-1);
		\draw [edge] (1,-1) -- (1.075,-1.5);
		\draw [edge] (1,-1) -- (1.025,-1.5);
		\draw [edge] (1,-1) -- (0.975,-1.5);
		\draw [edge] (1,-1) -- (0.925,-1.5);
	\draw [edge] (1,-0.5) -- (0.75,-1);
		\draw [edge] (0.75,-1) -- (0.775,-1.5);
		\draw [edge] (0.75,-1) -- (0.725,-1.5);
	\draw [edge] (1,-0.5) -- (0.5,-1); 
		\draw [edge] (0.5,-1) -- (0.55,-1.5); 
		\draw [edge] (0.5,-1) -- (0.5,-1.5);
		\draw [edge] (0.5,-1) -- (0.45,-1.5);
\draw [edge] (0,0) -- (0,-0.5);
	\draw [edge] (0,-0.5) -- (0.125,-1);
		\draw [edge] (0.125,-1) -- (0.2,-1.5);
		\draw [edge] (0.125,-1) -- (0.15,-1.5);
		\draw [edge] (0.125,-1) -- (0.10,-1.5);
		\draw [edge] (0.125,-1) -- (0.05,-1.5);
	\draw [edge] (0,-0.5) -- (-0.125,-1);
		\draw [edge] (-0.125,-1) -- (-0.175,-1.5);
		\draw [edge] (-0.125,-1) -- (-0.125,-1.5);
		\draw [edge] (-0.125,-1) -- (-0.075,-1.5);
\draw [edge] (0,0) -- (-1,-0.5);
	\draw [edge] (-1,-0.5) -- (-0.75,-1);
		\draw [edge] (-0.75,-1) -- (-0.85,-1.5);
		\draw [edge] (-0.75,-1) -- (-0.8,-1.5);
		\draw [edge] (-0.75,-1) -- (-0.75,-1.5);
		\draw [edge] (-0.75,-1) -- (-0.7,-1.5);
		\draw [edge] (-0.75,-1) -- (-0.65,-1.5);
	\draw [edge] (-1,-0.5) -- (-1,-1);
		\draw [edge] (-1,-1) -- (-1.025,-1.5);
		\draw [edge] (-1,-1) -- (-0.975,-1.5);
	\draw [edge] (-1,-0.5) -- (-1.25,-1);
		\draw [edge] (-1.25,-1) -- (-1.3,-1.5);
		\draw [edge] (-1.25,-1) -- (-1.25,-1.5);
		\draw [edge] (-1.25,-1) -- (-1.2,-1.5);
	\node[right] at (0.05,0) {level $ 3 $};
	\node[right] at (1.05,-0.5) {level $ 2 $};
	\node[right] at (1.5,-1) {level $ 1 $};
	\node[right] at (1.55,-1.5) {level $ 0 $};
\end{tikzpicture}
	\caption{ A sample of a Poisson GREM with $\tau =3$. There is a single vertex at level $\tau$ (the {``}top{''} of the tree). The paths from the top of the tree to the leaf vertices at level $0$ represent the configurations of the Poisson GREM. The energy of each configuration is given by the sum of the energies on the edges in the path.}
	\label{fig:3levels}
\end{figure}
	
	The model is defined on a tree of height $\tau$ with a single vertex at the top in a similar way to the standard GREM. The configurations are represented by the paths from the top of the tree to each of the leaf nodes at the bottom. An energy $E_b$ is associated with each edge $b$ in the tree and the energy of configuration $\mathcal{C}$ is given by the sum of the energies of all the bonds visited by the path associated with $\mathcal{C}$,
	
	\begin{equation}
	E_{\mathcal{C}}(\tau )=\sum _{b \in \text{  }\text{Path } \mathcal{C}} E_b.
	\end{equation}
	The Poisson GREM is constructed using a Poisson process with intensity $\rho _s(E)$ at each level $s$, $1\leq s\leq \tau$. In contrast to the standard GREM, the resulting tree has a varying branching rate as shown in the example in figure~\ref{fig:3levels}. 
	
	The method of construction is recursive starting from the top of the tree. The edges connecting the top level (level $\tau$) to level $\tau -1$ are constructed using the points of a Poisson process with intensity $\rho_{\tau }(E)$. An edge is constructed for each point $E_i$ and assigned energy $E_i$. These edges therefore connect the single level $\tau$ vertex to all the level $\tau-1$ vertices. This process is repeated for each of the vertices at level $\tau-1$  using independent Poisson point processes of intensity $\rho _{\tau-1}(E)$ to construct the level $\tau -2$ vertices and so on until one stops with the vertices at level $0$. An example of a tree resulting from this construction is given in Figure~\ref{fig:3levels}.
	
	The partition function for a particular sample of the Poisson GREM is given by
	
	\begin{equation}
	Z_{\tau }=\sum _{\mathcal{C}} \exp \left(-\beta  E_{\mathcal{C}}(\tau )\right)
	\end{equation}
	where $\beta$ is the inverse temperature and the sum is over all configurations, \( \mathcal{C} \), for the given sample. The Boltzmann weight of a given configuration $\mathcal{C}$
	is { }
	
	\begin{equation}
	W_{\mathcal{C}}=\frac{\exp \left(-\beta  E_{\mathcal{C}}(\tau )\right)}{Z_{\tau }}.
	\end{equation}
	
	The behaviour of the Poisson GREM depends on the choice of the Poisson intensity $\rho _s(E)$ used at each level $s=1,2,\ldots ,\tau$ of the tree. Here we take
	
	\begin{equation}
	\label{eq:ModelDefinition:rho=C_exp_AE-eBE2}
	\rho _s(E)=C_se^{A_s E-\epsilon  B_s E^2}
	\end{equation}
	where $A_s,B_s,C_s$ are positive constants and $ \epsilon $ is a small parameter. With suitable choices of these parameters we can map the Poisson GREM onto the standard GREM (see appendix \ref{sec:standard-vs-poisson-grem}). With this identification the small parameter $\epsilon =\frac{1}{N}$ where $N$ is the size of the system and the thermodynamic limit corresponds to $\epsilon \to 0$. As in the standard GREM there is a low temperature phase in which replica symmetry is broken \cite{Derrida_1986_Solution,Derrida_1986_Magnetic}. 
	
	In the following we will study two choices for the parameters in equation~\eqref{eq:ModelDefinition:rho=C_exp_AE-eBE2}.  First we consider is the gradual freezing case where $ A_1 > A_2>\cdots>A_\tau $. In this case we get a sequence of transitions as each successive level of the GREM freezes leading to full replica symmetry breaking when the number of levels become large. In the second case we will take the $ A_s=A $ independent of $ s $. This leads to a single phase transition where all levels of the GREM "freeze" simultaneously and we will refer to this as the simultaneous freezing case.
	
	The thermodynamic properties and finite size corrections to the free energy in both cases were analysed in~\cite{Cook_1991_Finite} in the context of the standard GREM. In the current paper we extend the approach to compute finite size corrections to overlap probabilities, which are central quantities in the replica theory.
		
\section{Recursion relations and the distribution of overlaps}
\label{sec:Recursion}	
	The analytic approach we use exploits the recursive nature of the Poisson GREM structure. The partition function of a particular sample can be expressed
	as
	
	\begin{equation}
	\label{eq:General:Recursion:Ztau=sum_exp_Ztau-1}
	Z_{\tau }=\sum \exp \left(-\beta  E_i\right) Z_{\tau -1}(i)
	\end{equation}
	where $\left\{E_1,E_2, \ldots \right\}$ are the points of the Poisson process at level $\tau$ for the given sample and $Z_{\tau -1}(i)$ is
	the partition function for the subtree connected to the $i$th edge at level $\tau$. What makes the GREM soluble is that the $ Z_{\tau -1}(i) $ are independent.
	
	We introduce the generating function 
	
	\begin{equation}
	\label{eq:General:Definition:Gtau=average_exp_exp_Z}
	G_{\tau }(x)=\left\langle \exp \left\{-e^{-\beta  x}Z_{\tau }(\beta )\right\}\right\rangle
	\end{equation}
	where the $\langle \ldots \rangle$ indicate an average over all the levels in the tree. It is clear that $0 \le G_\tau(x) \le 1$ and is monotonically increasing from zero for $ x $ large and negative to unity for $ x $ large and positive. It is useful to think of $ G_\tau $ as a wave front \cite{Derrida_1988_Polymers}.  In appendix~\ref{sec:appendix_recursion} we derive recursion relations that we summarise in this section.  The recursion for the partition function, equation~\eqref{eq:General:Recursion:Ztau=sum_exp_Ztau-1}, can be used to
	obtain a recursion for the generating function (see equation~\eqref{eq:Derivation:Recursion:Gtau=exp_integral_rho_Gtau-1}),
	
	\begin{equation}
	\label{eq:General:Recursion:Gtau=exp_integral_rho_Gtau-1}
	G_{s }(x)=\exp \left\{\int_{-\infty }^{\infty } \rho_{s }(E)\left[G_{s -1}(x+E)-1\right] \, \dd E\right\},
	\end{equation}
	with the initial condition (see equation~\eqref{eq:Derivation:InitialCondition:G0=exp_exp_beta_x}), 
	
	\begin{equation}
	\label{eq:General:InitialCondition:G0=exp_exp_beta_x}
	G_0(x)=\exp \left\{-e^{-\beta  x}\right\}.
	\end{equation}
	The recursion can be thought of as a travelling wave in discrete time \cite{Derrida_1988_Polymers,Cook_1991_Finite}. $ G_s(x) $ describes a wave front that moves a certain distance as we move up the tree. The temperature only enters through the initial condition and determines the initial shape of the wave front. The phase transitions we discuss later are closely related to the velocity selection mechanism observed for travelling waves \cite{Derrida_1988_Polymers}.
	This recursion was obtained as an approximation to the standard GREM in \cite{Cook_1991_Finite}. The thermodynamic limit of the quenched free energy and its finite size corrections, for example, can be obtained using an integral representation of the logarithm 
	
	\begin{equation}
	-\beta \langle F\rangle =\left\langle \log Z_{\tau }(\beta )\right\rangle =\beta \int _{-\infty }^{\infty }\left(G_0(x)-G_{\tau }(x)\right)\dd x.
	\end{equation}
	However, $G_{\tau }(x)$ is not sufficient to obtain the overlap function or to  characterise the replica symmetry breaking. To do this we need to define
	an additional generating function.
	
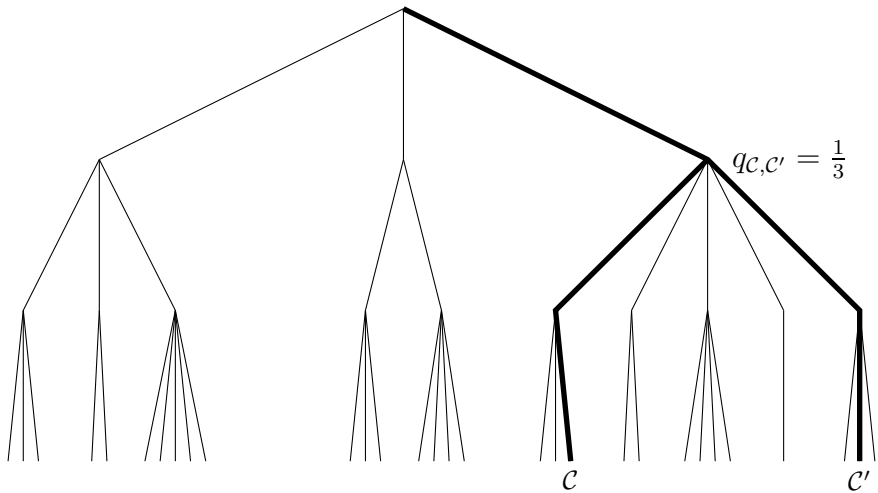
\begin{figure}
		\centering
\begin{tikzpicture}[scale=4, edge/.style={thin},thickEdge/.style={line width=2.0pt}]
\draw [edge] (0,0) -- (1,-0.5); 
	\draw [edge] (1,-0.5) -- (1.5,-1); 
		\draw [edge] (1.5,-1) -- (1.55,-1.5); 
		\draw [edge] (1.5,-1) -- (1.5,-1.5) node [below] {$ \mathcal{C'} $};
		\draw [edge] (1.5,-1) -- (1.45,-1.5);
	\draw [edge] (1,-0.5) -- (1.25,-1);
		\draw [edge] (1.25,-1) -- (1.25,-1.5);
	\draw [edge] (1,-0.5) -- (1,-1);
		\draw [edge] (1,-1) -- (1.075,-1.5);
		\draw [edge] (1,-1) -- (1.025,-1.5);
		\draw [edge] (1,-1) -- (0.975,-1.5);
		\draw [edge] (1,-1) -- (0.925,-1.5);
	\draw [edge] (1,-0.5) -- (0.75,-1);
		\draw [edge] (0.75,-1) -- (0.775,-1.5);
		\draw [edge] (0.75,-1) -- (0.725,-1.5);
	\draw [edge] (1,-0.5) -- (0.5,-1); 
		\draw [edge] (0.5,-1) -- (0.55,-1.5) node [below] {$ \mathcal{C} $}; 
		\draw [edge] (0.5,-1) -- (0.5,-1.5);
		\draw [edge] (0.5,-1) -- (0.45,-1.5);
\draw [edge] (0,0) -- (0,-0.5);
	\draw [edge] (0,-0.5) -- (0.125,-1);
		\draw [edge] (0.125,-1) -- (0.2,-1.5);
		\draw [edge] (0.125,-1) -- (0.15,-1.5);
		\draw [edge] (0.125,-1) -- (0.10,-1.5);
		\draw [edge] (0.125,-1) -- (0.05,-1.5);
	\draw [edge] (0,-0.5) -- (-0.125,-1);
		\draw [edge] (-0.125,-1) -- (-0.175,-1.5);
		\draw [edge] (-0.125,-1) -- (-0.125,-1.5);
		\draw [edge] (-0.125,-1) -- (-0.075,-1.5);
\draw [edge] (0,0) -- (-1,-0.5);
	\draw [edge] (-1,-0.5) -- (-0.75,-1);
		\draw [edge] (-0.75,-1) -- (-0.85,-1.5);
		\draw [edge] (-0.75,-1) -- (-0.8,-1.5);
		\draw [edge] (-0.75,-1) -- (-0.75,-1.5);
		\draw [edge] (-0.75,-1) -- (-0.7,-1.5);
		\draw [edge] (-0.75,-1) -- (-0.65,-1.5);
	\draw [edge] (-1,-0.5) -- (-1,-1);
		\draw [edge] (-1,-1) -- (-1.025,-1.5);
		\draw [edge] (-1,-1) -- (-0.975,-1.5);
	\draw [edge] (-1,-0.5) -- (-1.25,-1);
		\draw [edge] (-1.25,-1) -- (-1.3,-1.5);
		\draw [edge] (-1.25,-1) -- (-1.25,-1.5);
		\draw [edge] (-1.25,-1) -- (-1.2,-1.5);
\draw [thickEdge] (0,0) -- (1,-0.5) -- (1.5,-1) -- (1.5,-1.5); 
\draw [thickEdge] (1,-0.5) -- (0.5,-1) -- (0.55,-1.5);
	\node[right] at (1.05,-0.5) {{\large $ q_{\mathcal{C},\mathcal{C}'}=\frac{1}{3}$}};
\end{tikzpicture}
	\caption{An example of the overlap of two configurations for the 3 level Poisson GREM sample of figure~\ref{fig:3levels}. The paths from the top of the tree to configurations $ \mathcal{C} $ and $ \mathcal{C'} $ are indicated by thicker lines. In the example shown the two paths indicated have $r=1$ edges in common so that the overlap defined in~\eqref{eq:General:Overlap} is $q_{\mathcal{C},\mathcal{C}'}=\frac{1}{3 } $. }
		\label{fig:overlap}
	\end{figure}

	We first recall the ideas of overlap developed in the context of spin glasses \cite{Mezard_1987_Spin} and applied to the GREM \cite{Derrida_1986_Magnetic}. In the Poisson GREM the overlap between two configurations $\mathcal{C}$ and $\mathcal{C}'$ is the fraction
	of their length that the two paths share (see figure~\ref{fig:overlap}) . So that 
	
	\begin{equation} \label{eq:General:Overlap}
	q_{\mathcal{C},\mathcal{C}'}=\frac{r}{\tau }
	\end{equation}
	where the two configurations have $r$ edges in common, in particular $q_{\mathcal{C},\mathcal{C}}=1$. With this definition the distribution of overlaps $P(q)$ is then defined by
		
\begin{equation}
\label{eq:General:Definition:Pq=average_sum_WW_delta}
P(q) = \left\langle  \sum_{\mathcal{C},\mathcal{C}'} W_\mathcal{C} \: W_\mathcal{C'} \; \delta(q_{\mathcal{C},\mathcal{C}'} -q) \right\rangle
\end{equation}
where $\langle . \rangle$ denotes an average over all the random energies $E_b$ (i.e. an average over quenched disorder).

We have found that with the Poisson GREM the most effective way to approach the calculation of quantities like $ P(q) $ is to consider the probability $ Q_\tau^{(k)}(\tau-p+1) $ that $k$ copies of the same sample are in configurations that follow the same path from the top of the tree to a given level $p$ with no restriction on the path they follow below level $ p $. This means that they will be in configurations that have the same edges at level $\tau ,$ level $\tau -1$, ... , and level $p$ or equivalently they have the first $r=\tau-p+1$ edges in common starting from the root. Clearly from the definition 
\begin{equation}
 Q_\tau^{(k)}(0)=1 . 
\end{equation}
We show in appendix~\ref{sec:appendix_recursion} that the disorder average of this probability can be expressed as (for $ p \le \tau $)

\begin{equation}
\label{eq:General:Q=integral_of_H}
Q_\tau^{(k)}(\tau-p+1)=\frac{\beta }{\Gamma (k)}\int_{-\infty }^{\infty } H_\tau^{(p)}(x) \dd x
\end{equation}
where the generating function $H_s^{(p)}(x)$ is given by recursive integral equation.

\begin{equation}
	\label{eq:General:Recursion:Htau=Gtau_integral_rho_Htau-1}
H_s^{(p)}(x)= G_s(x)\int _{-\infty }^{\infty }\rho_s(E) \; H_{s-1}^{(p)}(x+E) \dd E
\end{equation}
with $ p+1 \le s \le \tau $ and the initial condition 

\begin{align}
	\label{eq:General:InitialCondition:Hpp=Gp_integral_rho_Gp-1}
H_p^{(p)}(x) & = G_p(x)e^{-k \beta  x }\int _{-\infty }^{\infty }\rho_p(E) \left[\frac{e^{\beta  x }}{\beta }\frac{\dd}{ \dd x}\right]^k \left( G_{p-1}(x+E)-1 \right) \dd E, \nonumber \\
  & = G_p(x)e^{-k \beta  x } \left[\frac{e^{\beta  x }}{\beta }\frac{\dd}{\dd x}\right]^k \ln G_{p}(x).
\end{align}
In the context of the travelling wave front $ H_s^{(p)}(x) $ looks like a travelling wave concentrated around the wave front and decaying to zero to the far left and far right of the wave front (in fact we will see that $ H_s^{(p)}(x) $ is closely related to $ G_s'(x) $). 

To obtain the distribution of overlaps \eqref{eq:General:Definition:Pq=average_sum_WW_delta} we only need $ Q_\tau^{(k)}(\tau-p+1) $ for $ k=2 $. The probability that two copies of a sample have precisely the first $ r $ bonds in common is given by $ Q_\tau^{(2)}(r)-Q_\tau^{(2)}(r+1) $. To obtain the continuous form of $ P(q) $ in the limit $ \tau \to \infty $ we note that we have 

\begin{equation}
	\label{eq:General:Q_to_P}
	Q_\tau^{(2)}(r)-Q_\tau^{(2)}(r+1) \to P(q) \dd q.
\end{equation}
with the overlap given by $ q=\frac{r}{\tau} $ and $ \frac{1}{\tau} \to \dd q $.

As we compute $ Q_\tau^{(k)}(r) $ for $ k>2 $ we can also look at more detailed aspects of the replica symmetry breaking (see \cite{Mezard_1987_Spin}). Within Parisi's replica symmetry breaking ansatz \cite{Parisi_1980_sequence,Mezard_1984_Nature,Mezard_1984_Replica,Derrida_1997_random} one expects that
\begin{equation} \label{eq:General:ReplicaOverlap}
	Q_\tau^{(k)}(r)=
		\lim\limits_{n \to 0} 
			\frac{n(\mu - 1)(\mu -2) \cdots (\mu -k+1)}
				{n(n-1)(n-2)\cdots(n-k+1)}
		=\frac{\Gamma(k-\mu)}{\Gamma(k)\Gamma(1-\mu)}
\end{equation}
where $ \mu $ is the number of replicas in a block at an overlap of $ q=\frac{r}{\tau} $. (This is simply the number of ways of choosing $ k $ different replicas among $ n $ in such a way that they all belong to the same block). In the Parisi theory, together with $ n \to 0 $ limit, the block size $ \mu $ becomes a real function on the interval $ [0,1] $ that is determined by the saddle point conditions. We can eliminate $ \mu $ from \eqref{eq:General:ReplicaOverlap} to obtain the relationship (\ref{eq:General:Yk_Fk},~\ref{eq:General:Fk_definition}) between the overlap probabilities 
\begin{equation}
	\label{eq:General:Q_relationship}
	Q_\tau^{(k)}(r)
	=\frac{\Gamma(k-1+Q_\tau^{(2)}(r))}{\Gamma(k) \: \Gamma(Q_\tau^{(2)}(r))}
\end{equation}
which is therefore a direct consequence of the Parisi ansatz. We are going to see in section~\ref{sec:Gradual} how this relationship is modified by finite size effects. 

Remark: Taking the derivative of equation~\eqref{eq:General:Recursion:Gtau=exp_integral_rho_Gtau-1} we obtain a recursion for $G_s^{\prime}(x)$

\begin{equation}
\label{eq:Gdashed}
G_s'(x)=G_s(x)\int _{-\infty }^{\infty }\rho_s(E-x)\,G_{s-1}^{\prime }(E) \dd E.
\end{equation}
Comparing with formula~\eqref{eq:General:Recursion:Htau=Gtau_integral_rho_Htau-1} we see that $G_s^{\prime }(x)$ is a solution to the recursion satisfied by $H_s^{(p)}(x)$. However it does not necessarily satisfy the initial condition~\eqref{eq:General:InitialCondition:Hpp=Gp_integral_rho_Gp-1}. We will see below that under iteration $ H_s^{(p)}(x) $ becomes, after one step, proportional to $ G_s^{\prime}(x) $ with a pre-factor independent of $x$. This will greatly simplify the calculation and in addition the integration in \eqref{eq:General:Q=integral_of_H} becomes trivial.

\section{Gradual freezing case} \label{sec:Gradual}
 The first case we consider is the gradual freezing case, where we assume in equation~\eqref{eq:ModelDefinition:rho=C_exp_AE-eBE2} that $ A_1 > A_2 > \cdots > A_\tau $. In this case we get a sequence of transitions as each successive level of the GREM freezes. We will assume that $ \beta > A_1 $ so that we are in the lowest temperature phase.
	
\subsection{Recursion on $G_s(x)$}

It turns out that,  under the recursion~\eqref{eq:General:Recursion:Gtau=exp_integral_rho_Gtau-1}, one can write $G_s(x)$ at order $1$ in $\epsilon$ as
\begin{equation}
\label{eq:Gradual:G_solution}
G_s(x)=\exp \left\{-e^{-A_s\left(x-\mu _s\right)}\right\} \Big(1 + \epsilon \, e^{-A_s(x-\mu_s)}\,  g_s(\mu_s-x) + O(\epsilon^2) \Big)
\end{equation}
where $g_s(x)$ is a polynomial of degree $2$ in $x$. As we will now see, this can be checked by recursion. Using the fact that (for $0<A_s<A_{s-1}$ and  for small enough $\gamma$, i.e. $0 < \gamma < A_{s-1}-A_s$ ) one can show that  the following integrals $I_s(\gamma)$ and $J_s(\gamma)$ are given by
\begin{align} \label{eq:Gradual:I_integral}
I_s(\gamma) \equiv & \;
C_s \int_{-\infty}^{\infty} \dd E \  e^{(A_s + \gamma) E} \  
\left[ \exp \left\{-e^{-A_{s-1}\left(E+ x-\mu_{s-1} \right)}\right\}  -1 \right] \nonumber
\\  \ = &  \
{C_s \over A_{s-1}} \, \Gamma \left(-{\gamma+ A_s \over A_{s-1}} \right) \, e^{-(A_s+ \gamma) (x-\mu_{s-1})} 
\end{align}

\begin{align}
J_s(\gamma) \equiv & \;
C_s \int_{-\infty}^{\infty} \dd E \  e^{A_s E} \  
\exp \left\{-e^{-A_{s-1}\left(E+ x-\mu_{s-1} \right)}\right\}   e^{-(A_{s-1}+\gamma) (E+x-\mu_{s-1})} \nonumber
\\  \ = &  \
{C_s \over A_{s-1}} \, \Gamma \left(1+ {\gamma -A_s \over A_{s-1}} \right) \, e^{-A_s(x-\mu_{s-1})} 
\end{align}
One can check that
$$J_s(0)=- {A_s \over A_{s-1}} \, I_s(0)$$
and  it is easy to see that at order  $1$ in $\epsilon$ one has
\begin{multline}
\int_{-\infty }^{\infty } \rho_{s }(E)  \left[G_{s -1}(E+x)-1\right] \, \dd E =  I_s(0) + \epsilon \left[ - B_s I_s''(0) \vphantom{{g_{s-1}''(0) \over 2}} \right.  \\
 \left. {} + g_{s-1}(0) J_s(0) + g_{s-1}'(0) J_s'(0) +{g_{s-1}''(0) \over 2} J_s''(0) \right] +  O(\epsilon^2) \nonumber .
\end{multline}
The ratios
$J_s'(0)/J_s(0),J_s''(0)/J_s(0)$ are independent of $x$ whereas the ratio $I_s''(0)/I_s(0)$ 
is quadratic in $x$.
Then using (\ref{eq:General:Recursion:Gtau=exp_integral_rho_Gtau-1})
one can see that the form~\eqref{eq:Gradual:G_solution} is stable and that $\mu_s$ and $g_s(x)$ are given by

\begin{equation} \label{eq:Gradual:mu_recursion}
e^{A_s \mu_s} = - 
{C_s \over A_{s-1}} \, \Gamma \left(- {A_s \over A_{s-1}} \right) \, e^{A_s\mu_{s-1}}
\end{equation}
and
\begin{align} \label{eq:Gradual:g_result}
	g_s(\mu_s-x) =  - \frac{1}{I_s(0)} & \left[ - B_s I_s''(0)  +  g_{s-1}(0) J_s(0) \right. \nonumber \\ 
	& \qquad \left. {} + g_{s-1}'(0) J_s'(0) + \frac{g_{s-1}''(0)}{2} J_s''(0) \right] 
\end{align}
so that $g_s(\mu_s-x)$ is indeed quadratic in $x$.

\subsection{Recursion on  $H_s^{(p)}(x)$}

We now need to find the form of the generating function $H_s^{(p)}(x)$ that satisfies the recursion~\eqref{eq:General:Recursion:Htau=Gtau_integral_rho_Htau-1} and the initial condition~\eqref{eq:General:InitialCondition:Hpp=Gp_integral_rho_Gp-1} for small $ \epsilon $. We start with the initial condition and use the form~\eqref{eq:Gradual:G_solution} for $ G_s(x) $. Using recursion over $k$ it is then reasonably easy to show that

\begin{multline*}
	e^{-k \beta  x }\left(\frac{e^{\beta  x }}{\beta }\frac{\dd}{\dd x}\right)^k
	\log [G_p(x)] = - e^{-A_p(x-\mu_p)} \  { \times} 
	\left[ {\Gamma\left(k +z
		\right) \over \Gamma\left(z \right)} (1- \epsilon g_p(\mu_p-x)) \right.
	\\ \left. \left. + {\epsilon \over \beta} \left({\Gamma\left(k +z \right) \over \Gamma\left(z \right)} \right)' g_p'(\mu_p-x)
	-  {\epsilon \over 2 \beta^2}  \left({\Gamma\left(k +z \right) \over \Gamma\left(z \right)} \right)'' g_p''(\mu_p-x)
	\right] 
	\right|_{z=- {A_p \over \beta}}
\end{multline*}
So that the initial condition (\ref{eq:General:InitialCondition:Hpp=Gp_integral_rho_Gp-1}) gives to $ O(\epsilon) $
\begin{multline*}
H_p^{(p)}(x)=   - 
\exp \left\{-e^{-A_p\left(x-\mu_p\right)}\right\}
\Bigg[  e^{-A_p(x-\mu_p)}  \Bigg( {\Gamma\left(k +z
	\right) \over \Gamma\left(z \right)} (1- \epsilon g_p(\mu_p-x))
\\ + {\epsilon \over \beta} \left({\Gamma\left(k +z \right) \over \Gamma\left(z \right)} \right)' g_p'(\mu_p-x)
-  {\epsilon \over 2 \beta^2}  \left({\Gamma\left(k +z \right) \over \Gamma\left(z \right)} \right)'' g_p''(\mu_p-x)
\Bigg)
\\ \left. + \epsilon\  {\Gamma\left(k +z
	\right) \over \Gamma\left(z \right)} \  e^{-2 A_p(x-\mu_p)}   g_p(\mu_p-x) + O(\epsilon^2) \Bigg]
\right|_{z=- {A_p \over \beta}}
\end{multline*}
Using the derivative of equation~\eqref{eq:Gradual:G_solution} one has
\begin{multline*}
G_p'(x)= \exp \left\{-e^{-A_p\left(x-\mu_p\right)}\right\}
\Bigg[  e^{-A_p(x-\mu_p)}   \Big(
A_p- \epsilon A_p g_p(\mu_p-x)
\\ - \epsilon   g_p'(\mu_p-x)
\Big)
 + \epsilon A_p  e^{-2 A_p(x-\mu_p)}   g_p(\mu-x) \Bigg].
\end{multline*}
So one can write 
$$
H_p^{(p)}(x)= - {1 \over A_p}  {\Gamma\left(k  -{A_p\over \beta}
	\right) \over \Gamma\left(  -{A_p\over \beta}\right)}  G_p'(x)  + \epsilon 
\exp \left\{-e^{-A_p\left(x-\mu_p\right)}\right\}
e^{-A_p(x-\mu_p)}  \ h_p(x)
$$
where
\begin{multline} \label{eq:Gradual:small_h}
h_p(x) = 
\Bigg[
- {1\over A_p} g_p'(\mu_p-x)
{\Gamma\left(k +z \right) \over \Gamma\left(z \right)} 
-{1 \over \beta }g_p'(\mu_p-x) \left({\Gamma\left(k +z \right) \over \Gamma\left(z \right)} \right)'
\\ \left. {}+ \frac{1}{2 \beta^2} g_p''(\mu_p-x)  \left({\Gamma\left(k +z \right) \over \Gamma\left(z \right)} \right)'' \Bigg] \right|_{z=- {A_p \over \beta}}
\end{multline}
Then substituting into the recursion~\eqref{eq:General:Recursion:Htau=Gtau_integral_rho_Htau-1}, at order $\epsilon$, we obtain 
\begin{multline} \label{eq:Gradual:Hp_plus_1}
H_{p+1}^{(p)}(x)=G_{p+1}'(x) 
	\Bigg[\frac{\Gamma \left(k-\frac{A_p}{\beta}\right) }{\beta\Gamma
		\left(1-\frac{A_p}{\beta}\right)} 
		+ \epsilon \frac{h_p(\mu_p)}{A_p}
		\\ - \epsilon \frac{h_p'(\mu_p)}{A_p^2} \frac{\Gamma' \left(1- \frac{A_{p+1}}{A_p}\right)}{\Gamma \left(1- \frac{A_{p+1}}{A_p}\right)}
		+ O\left( \epsilon^2 \right)
	\Bigg]
\end{multline}

It is clear (see \eqref{eq:General:Recursion:Htau=Gtau_integral_rho_Htau-1} and \eqref{eq:Gdashed}) that the form \eqref{eq:Gradual:Hp_plus_1} satisfies the recursion \eqref{eq:General:Recursion:Htau=Gtau_integral_rho_Htau-1} and we can compute the overlap probability using \eqref{eq:General:Q=integral_of_H}. 

\begin{multline}
Q_\tau^{(k)}(\tau-p+1)= \frac{\beta}{\Gamma(k)}
\Bigg[\frac{\Gamma \left(k-\frac{A_p}{\beta}\right) }{\beta\Gamma
	\left(1-\frac{A_p}{\beta}\right)} 
+ \epsilon \frac{h_p(\mu_p)}{A_p}
\\ - \epsilon \frac{h_p'(\mu_p)}{A_p^2} \frac{\Gamma' \left(1- \frac{A_{p+1}}{A_p}\right)}{\Gamma \left(1- \frac{A_{p+1}}{A_p}\right)}
+ O\left( \epsilon^2 \right)
\Bigg]	
\end{multline}
We can use \eqref{eq:Gradual:I_integral}, \eqref{eq:Gradual:mu_recursion}, \eqref{eq:Gradual:g_result} and \eqref{eq:Gradual:small_h} to obtain
$$g_p'(0) = 2 B_p (\mu_{p-1} - \mu_p)  - 2 \frac{B_p}{A_{p-1}} {
	\Gamma'(-\frac{A_p}{A_{p-1}})
	\over \Gamma(- \frac{A_p}{A_{p-1}}  ) } $$
$$g_p''(0) = 2 B_p $$
and after some algebra put the overlap probability in the form
\begin{equation} \label{eq:Gradual:Q_first_order}
Q_\tau^{(k)}(\tau-p+1)= \left.
\left[1+ \epsilon \Delta_1 \frac{\dd}{\dd z}
		+\epsilon \Delta_2 \frac{\dd^2}{\dd z^2} + O\left( \epsilon^2 \right)
\right] 
\frac{\Gamma(k+z)}{\Gamma(k)\Gamma(1+z)} \right|_{z=-\frac{A_p}{\beta}}
\end{equation}
where 
\begin{equation} \label{eq:Gradual:Delta_1}
\Delta_1= \frac{2B_p}{\beta} \left[
	\mu_{p-1}
	-\frac{\Gamma' \left(1- \frac{A_p}{A_{p-1}}\right)}{A_{p-1}\Gamma \left(1- \frac{A_{p}}{A_{p-1}}\right)}
	-\mu_p
	+\frac{\Gamma' \left(1- \frac{A_{p+1}}{A_p}\right)}{A_p \Gamma \left(1- \frac{A_{p+1}}{A_p}\right)}
\right]
\end{equation}
and
\begin{equation} \label{eq:Gradual:Delta_2}
\Delta_2=- \frac{B_p}{\beta^2}.
\end{equation}
The expression \eqref{eq:Gradual:Q_first_order} suggests that the order $ \epsilon $ correction can be viewed as a small shift in the transition temperature (through the parameter $ z $) from the $ \Delta_1 $ term and fluctuation from the $ \Delta_2 $ term. 

With a little more algebra we can transform \eqref{eq:Gradual:Q_first_order} so that, at order $ \epsilon $ and with $ k \ge 3 $,
\begin{equation} 
Q_\tau^{(k)}(r)= \left.
\left[1+ +\epsilon \Delta_2 \frac{\dd^2}{\dd z^2} + O\left( \epsilon^2 \right)
\right] 
\frac{\Gamma(k-1+z)}{\Gamma(k)\Gamma(z)} \right|_{z=Q_\tau^{(2)}(r)}
\end{equation}
which is the relation announced in~\eqref{eq:General:Yk_Fk}, \eqref{eq:General:Fk_definition} and \eqref{eq:General:Yk_finite_size} if one identifies $ Q_\tau^{(k)}(r) $ with $ \left\langle Y_k\right\rangle $ as they are the same quantity.

Remark: We would expect that in the case $ \tau = 1 $ these results should apply to the Poisson REM analysed in~\cite{Derrida_2015_Finite} and defined by parameters $ \alpha, \beta, A $. We do in fact recover equation (38) of~\cite{Derrida_2015_Finite} if we take $ A_0=\beta $, $ A_1=\alpha $, $ A_2=0 $, $ B_1=1 $ and $ C_1=A $ in equations~\eqref{eq:Gradual:Q_first_order}, \eqref{eq:Gradual:Delta_1} and \eqref{eq:Gradual:Delta_2} and recognise that $ \left\langle P_k\right\rangle $ in \cite{Derrida_2015_Finite} is the same as $ Q_1^{(k)}(1) $ in the current paper.

\section{Simultaneous freezing case}
In the simultaneous freezing case we will take the $ A_s=A $ independent of $ s $ in equation~\eqref{eq:ModelDefinition:rho=C_exp_AE-eBE2}. This leads to a single phase transition where all levels of the GREM "freeze" simultaneously. The transition occurs when the inverse temperature $ \beta = A $. We will consider the behaviour at the transition temperature, $ \beta=A $, and in the low temperature phase when  $ \beta > A $.

\subsection{The solution for $G_s(x)$}
In the simultaneous freezing case we expect $ G_s(x) $ to have a similar travelling wave form to the gradual freezing case \cite{Cook_1991_Finite} . However, one needs to know the form of $ G_s(x) $ in a larger region of $ O\left( \epsilon^{-1/2}\right) $ around the wave front. The reason is that if we simply expand the disorder distribution in equation~\eqref{eq:ModelDefinition:rho=C_exp_AE-eBE2} in small $\epsilon$ (as we did in the gradual freezing case) the integrals resulting from the recursion~\eqref{eq:General:Recursion:Gtau=exp_integral_rho_Gtau-1} diverge when $ A_s=A_{s-1} $ and $ \epsilon \rightarrow 0 $. As a consequence we must keep the full Gaussian form of the disorder distribution in equation~\eqref{eq:ModelDefinition:rho=C_exp_AE-eBE2}.
If one assumes $ G_s(x) $ has the form \cite{Cook_1991_Finite}
\begin{equation}
	\label{eq:Simultaneous:ZerothOrder:Gtau=exp_f_exp_Ax}
	G_s(x)=\exp \left\{-f_s\left((x-\mu_s)\sqrt{\epsilon }\right) e^{-A(x-\mu_s)}\right\}
\end{equation}
in the region where $ x-\mu_s= O\left( \epsilon^{-1/2}\right) $ and for small $\epsilon$, then the recursion~\eqref{eq:General:Recursion:Gtau=exp_integral_rho_Gtau-1} will give the same form for $ G_{s+1}(x) $. We require that $ f_s(0)= O(1) $ so that when $ x = \mu_s $ we have $ G(\mu_s) = \exp(f_s(0)) $ and therefore $ \mu_s $ gives an indication of the position of the wave front.

We find from the recursion~\eqref{eq:General:Recursion:Gtau=exp_integral_rho_Gtau-1} that for small $ \epsilon $ and $s \geq 2$ 
\begin{multline}
\label{eq:Simultaneous:Recursion:fs=integral}
f_s(v) \simeq \frac{C_s}{\sqrt{\epsilon }} e^{-A(\mu_s-\mu_{s-1})}
	\\ \times \int _{-\infty}^{\infty } e^{A \frac{w}{\sqrt{\epsilon}}}
	e^{-B_s(v-w)^2}\left[ 1- 
	\exp \left\lbrace -f_{s-1}(w) e^{-A \frac{w}{\sqrt{\epsilon}}}\right\rbrace 	
		\right] \dd w,
\end{multline}
where we have assumed that $ (\mu_s-\mu_{s-1}) $ is smaller than order $ \frac{1}{\sqrt{\epsilon}} $ and discarded lower order terms. We can now take

\begin{equation}
	\label{eq:Simultaneous:Recursion:mu}
	\mu_s= \mu_{s-1} + \frac{1}{A} \ln\left[ \frac{C_s}{\sqrt{\epsilon}}\right] .
\end{equation}
Now for $ \epsilon $ sufficiently small in \eqref{eq:Simultaneous:Recursion:fs=integral} we have
\begin{equation}
	\label{eq:Simultaneous:Recursion:ftau=integral_of_ftau-1}
f_s(v)=\int _0^{\infty }e^{-B_s(v-w)^2}f_{s-1}(w) \dd w.
\end{equation}
To obtain the initial condition for this pair of recursions we substitute $ G_0(x) $ from~\eqref{eq:General:InitialCondition:G0=exp_exp_beta_x} into the right hand side of recursion~\eqref{eq:General:Recursion:Gtau=exp_integral_rho_Gtau-1}. For small $ \epsilon $ we must distinguish the two cases $ \beta=A $ and $ \beta>A $.

When $\beta =A$ the analysis is similar to \eqref{eq:Simultaneous:Recursion:fs=integral} above; the Gaussian term in equation~\eqref{eq:ModelDefinition:rho=C_exp_AE-eBE2} is required
for convergence and as a consequence we find

\begin{equation}
\label{eq:Simultaneous:Recursion:mu1atTc}
\mu_1= \frac{1}{A} \ln\left[ \frac{C_1}{\sqrt{\epsilon}}\right] ,
\end{equation}

\begin{equation}
	\label{eq:Simultaneous:InitialCondition:f1=integral_exp}
f_1(v)=\int _0^{\infty }e^{-B_1(v-w)^2} \dd w.
\end{equation}
When $\beta >A$ we no longer require the Gaussian term for convergence and we obtain 

\begin{equation}
\label{eq:Simultaneous:Recursion:mu1belowTc}
\mu_1= \frac{1}{A} \ln\left[ \frac{C_1}{A}\Gamma \left(1-\frac{A}{\beta }\right)\right] ,
\end{equation}

\begin{equation}
	\label{eq:Simultaneous:InitialCondition:f1=gamma_function}
f_1(v)= e^{-B_1v^2}.
\end{equation}

\subsection{The solution for $H_s^{(p)}(x)$}
First we consider the initial condition~\eqref{eq:General:InitialCondition:Hpp=Gp_integral_rho_Gp-1} using the form~\eqref{eq:Simultaneous:ZerothOrder:Gtau=exp_f_exp_Ax}. The result is

\begin{equation}
	\label{eq:Simultaneous:InitialCondition:Hrr=-Gr_derivatives_fr}
H_p^{(p)}(x)=-G_p(x) e^{-\beta  k x}\left[ \frac{e^{\beta  x}}{\beta }\frac{\dd}{\dd x}\right]^kf_p\left((x-\mu_p)\sqrt{\epsilon }\right) e^{-A (x-\mu_p)}.
\end{equation}
the crucial observation is that each derivative of $f_p\left((x-\mu_p)\sqrt{\epsilon }\right)$ produces a factor of $\sqrt{\epsilon }$, so that to order $\sqrt{\epsilon }$ we can forget derivatives of $ f_p $ beyond $ f_p' $. 
One can show from \eqref{eq:Simultaneous:InitialCondition:Hrr=-Gr_derivatives_fr} that 
\begin{multline}
\label{eq:Simultaneous:Hpp-first}
H_p^{(p)}(x)= - G_p(x)\, e^{-A (x-\mu_p)}
\left[f_p\left((x-\mu_p)\sqrt{\epsilon}\right)  \left[\frac{\Gamma(k+z)}{\Gamma(z)}\right] \right.  \\
\left.
 \qquad { } \left. + \frac{\sqrt{\epsilon}}{\beta}f_p^{\prime }\left((x-\mu_p)\sqrt{\epsilon}\right)  \frac{\dd}{\dd z} \left[\frac{\Gamma(k+z)}{\Gamma(z)}\right]  + O(\epsilon)\right]
 \right|_{z=-\frac{A}{\beta}}.
\end{multline}
Using the form \eqref{eq:Simultaneous:ZerothOrder:Gtau=exp_f_exp_Ax} one has 
\begin{multline}
\label{eq:Simultaneous:Gprime}
G_p'(x)= G_p(x)\, e^{-A (x-\mu_p)}
\Big[A \, f_p\left((x-\mu_p)\sqrt{\epsilon}\right)  
 \\+ \sqrt{\epsilon} \, f_p^{\prime }\left((x-\mu_p)\sqrt{\epsilon}\right) + O(\epsilon)\Big]
\end{multline}
So that \eqref{eq:Simultaneous:Hpp-first} can be rewritten as
\begin{multline}
	\label{eq:Simultaneous:Solution:Hrr=hr0_Gprime+G_fprime}
H_p^{(p)}(x)= 
	\Bigg[G_p^{\prime }(x)+\sqrt{\epsilon} \,G_p(x)\text{  }e^{-A (x-\mu_p)}f_p^{\prime }\left((x-\mu_p)\sqrt{\epsilon}\right) \frac{A}{\beta}  \frac{\dd}{\dd z} 
	\\ + O(\epsilon)\Bigg] 
	\left. \left[\frac{\Gamma(k+z)}{\beta \, \Gamma(1+z)}\right] \right|_{z=-\frac{A}{\beta}}.
\end{multline}
One can then check using \eqref{eq:General:Recursion:Htau=Gtau_integral_rho_Htau-1} and \eqref{eq:Simultaneous:ZerothOrder:Gtau=exp_f_exp_Ax} that after iteration $ H_s^{(p)}(x) $ remains of the form

\begin{multline}
\label{eq:Simultaneous:Solution:Hsp}
H_s^{(p)}(x)= 
\Bigg[G_s^{\prime }(x)+\sqrt{\epsilon} \, G_s(x)\text{  }e^{-A (x-\mu_s)}h_s^{(p)}\left((x-\mu_s)\sqrt{\epsilon}\right) \frac{A}{\beta}  \frac{\dd}{\dd z} 
\\ {} + O(\epsilon)\Bigg] 
\left. \left[\frac{\Gamma(k+z)}{\beta \, \Gamma(1+z)}\right] \right|_{z=-\frac{A}{\beta}}
\end{multline}
where we define the function $h_s^{(p)}(v)$ for $s \geq p$ by the recursion 

\begin{equation}
	\label{eq:Simultaneous:Definition:htilde=integral_on_htilde-1}
	h_s^{(p)}(v)=\int _0^{\infty }e^{-B_s(v-w)^2}h_{s-1}^{(p)}(w) \dd w
\end{equation}
with the initial condition 

\begin{equation}
	\label{eq:Simultaneous:InitialCondition:htilde=fprime}
	h_p^{(p)}(v)=f_p^{\prime }(v) .
\end{equation}

\subsection{The overlap $ Q_\tau^{(k)}(r) $}
Substituting\eqref{eq:Simultaneous:Solution:Hsp} in~\eqref{eq:General:Q=integral_of_H}  gives the overlap probability 

\begin{equation}
\label{eq:Simultaneous:Q_first_correction}
Q_\tau^k(\tau-p+1)=\left[1+\sqrt{\epsilon} \,
	\frac{h_\tau^{(p)}(0)}{\beta f_\tau(0)} \frac{\dd}{\dd z} + O(\epsilon )\right]
	\left. \left[\frac{\Gamma(k+z)}{\Gamma(k) \Gamma(1+z)}\right] \right|_{z=-\frac{A}{\beta}}.
\end{equation}
We can interpret the finite size correction as a small shift in the transition temperature (more precisely it is a shift in the parameter $ z= -\frac{A}{\beta} $) If we define 
\begin{equation}
	\Delta z = \sqrt{\epsilon} \: \frac{h_p^{(p)}(0)}{\beta f_\tau(0)}
\end{equation}
then
\begin{equation}
	\label{eq:Simultaneous:Q_of_deltaz}
	Q_\tau^k(\tau-p+1)=\frac{\Gamma \left(k-\frac{A}{\beta }+ \Delta z \right)}{\Gamma (k)\Gamma \left(1-\frac{A}{\beta } + \Delta z \right)}
		+ O(\epsilon )
\end{equation}
and \eqref{eq:General:Q_relationship} is satisfied to order $ \sqrt{\epsilon} $. It would be interesting to see if, as in the gradual freezing case, the relationship is broken by the next correction (i.e at order $ \epsilon $) as in  (\ref{eq:General:Fk_definition},~\ref{eq:General:Yk_finite_size},~\ref{eq:General:Delta_2}).

\subsection{The function $ P(q) $ and the branching random walk}
In our earlier paper we computed the probability distribution of overlaps $ P(q) $ for a branching random walk by approximating it with a GREM \cite{Derrida_2016_Genealogy,Schmidt_2015_Derridas}. In this section we show how to recover this result using the current Poisson GREM model.
The difference between overlaps for the Poisson GREM from \eqref{eq:Simultaneous:Q_first_correction} is

\begin{multline} \label{eq:Simultaneous:Qk_difference_h}
Q_\tau^k(\tau-p+1)-Q_\tau^k(\tau-p+2)\\ 
= \sqrt{\epsilon}
	\frac{\left(h_\tau^{(p)}(0)-h_\tau^{(p-1)}(0)\right)}{\beta f_\tau(0)} \frac{\dd}{\dd z}
 	\left. \left[\frac{\Gamma(k+z)}{\Gamma(k) \Gamma(1+z)}\right] \right|_{z=-\frac{A}{\beta}}+ O(\epsilon ).
\end{multline}
We can define $\tilde{f}_s^{(p)}(v)$ such that
\begin{equation}
	h_s^{(p)}(v)-h_s^{(p-1)}(v)=\tilde{f}_s^{(p)}(v) f_{p-1}(0), \,  s\geq p.
\end{equation}
From \eqref{eq:Simultaneous:InitialCondition:htilde=fprime} after some integration by parts one can check that
\begin{equation}
\tilde{f}_p^{(p)}(v) =e^{-B_pv^2}.
\end{equation}
Also using the $h$ recursion~\eqref{eq:Simultaneous:Definition:htilde=integral_on_htilde-1} we can  thus obtain a recursion for $\tilde{f}_s^{(p)}(v)$ with $ s \ge p+1 $
\begin{equation}
\tilde{f}_s^{(p)}(v)\text{  }=\int _0^{\infty }e^{-B_s(v-w)^2}\tilde{f}_{s-1}^{(p)}(w) \dd w.
\end{equation}
Then \ref{eq:Simultaneous:Qk_difference_h} becomes

\begin{multline}
	Q_\tau^k(\tau-p+1)-Q_\tau^k(\tau-p+2)
	\\ =\sqrt{\epsilon} \,
	\frac{\tilde{f}_\tau^{(p)}(0)
		f_{\tau-p}(0)}{\beta f_\tau(0)} \frac{\dd}{\dd z}
	\left. \left[\frac{\Gamma(k+z)}{\Gamma(k) \Gamma(1+z)}\right] \right|_{z=-\frac{A}{\beta}}+ O(\epsilon ).
\end{multline}
Letting $r=\tau-p+1$ and $ k=2 $ then

\begin{equation}
Q_\tau^{(2)}(r)-Q_\tau^{(2)}(r+1)=\frac{\sqrt{\epsilon }}{\beta}\frac{\tilde{f}_\tau^{(\tau-r+1)}(0)
	f_{\tau-r}(0)}{f_\tau(0)} + O(\epsilon ).
\end{equation}
The integrals $ f $ and $ \tilde{f} $ can be computed (see (25) and (26) of \cite{Derrida_2016_Genealogy}) if we assume that $ B_s=B $ for all $ s $. If we also take $ \tau $, $ r $ and $ \tau-r \gg 1$ then we can use~\eqref{eq:General:Q_to_P} to obtain the probability distribution of overlaps. We find for $ \beta=A $
\begin{equation}
\label{eq:Simultaneous:P_at_Tc}
P(q)= \frac{1}{\beta} \sqrt{\frac{\epsilon B}{\pi \tau}} \frac{1}{q^{\frac{3}{2}}(1-q)^{\frac{1}{2}}} \qquad,
\end{equation}
and for $ \beta > A $
\begin{equation}
	\label{eq:Simultaneous:P_below_Tc}
	P(q)= \frac{1}{\beta} \sqrt{\frac{\epsilon B}{\pi \tau}} \frac{1}{(q(1-q))^{\frac{3}{2}}} \qquad .
\end{equation}
In order to map this to the binary branching random walk of height $ t $ we approximate $ n $ levels of the binary tree with a single level of the Poisson GREM so that $ t=n \tau $ as described in \cite{Derrida_2016_Genealogy}. The lowest energy levels of the binary tree are well approximated for $ n \gg 1 $ if we take
\begin{equation}
	\sqrt{\frac{\epsilon B}{\pi \tau}} \to \frac{1}{\sqrt{t}} \frac{1}{\sqrt{2 \pi \beta_c v''(\beta_c)}},
\end{equation}
in \eqref{eq:Simultaneous:P_below_Tc} and \eqref{eq:Simultaneous:P_at_Tc}.
It is interesting to note that in both the binary tree and the GREM finite size corrections are sufficient to produce full replica symmetry breaking in a model that has only one step of replica symmetry breaking in the thermodynamic limit.

(Note: In our branching random walk paper \cite{Derrida_2016_Genealogy} we incorrectly transcribed an equation. It does not affect the final results of the calculation or the argument, but equation (27) of \cite{Derrida_2016_Genealogy} should be 
\[ P(E=n\epsilon) \dd E \sim \sqrt{\frac{f''\left(\frac{E}{n}\right)}{2 \pi n}} \exp \left[-nf \left(\frac{E}{n}\right)\right]\dd E. \]
The pre-factor is incorrect in the published version. The equation also appears in the appendix and should be corrected there also.)

\section{Summary}
In this paper we have extended our earlier work \cite{Derrida_2015_Finite,Derrida_2016_Genealogy} on the effect of finite size corrections on replica symmetry breaking in simple models that can be successfully analysed without the replica method. Here we have studied two scenarios for the Poisson GREM; the gradual freezing case where there are a sequence of replica symmetry breaking steps and the simultaneous freezing case where these transitions coincide and there is a one step replica symmetry breaking transition. In both cases we focus on the low temperature phase.

In the gradual case the finite size effects are similar to the Poisson REM studied in \cite{Derrida_2015_Finite}. They can be interpreted as a small shift in the transition temperature and fluctuations in the block size used in Parisi's replica symmetry breaking scheme with, as in the REM \cite{Derrida_2015_Finite}, a negative variance. Here we also observe that one of the predictions of Parisi's scheme, equation \eqref{eq:General:Q_relationship}, has to be modified when we include the leading finite size correction. 

In the simultaneous freezing case we have shown that the leading finite size correction is enough to produce full replica symmetry breaking in a model which has a single step of replica symmetry breaking in the thermodynamic limit. At this order relations (\ref{eq:General:Yk_Fk},~\ref{eq:General:Fk_definition}) is satisfied. Whether a negative variance would appear at the next order is an open question. 

It would be interesting to look at models which do not have built in ultrametricity to see at what level of the finite size corrections the non-ultrametric structure becomes apparent~\cite{Bolthausen_2006_nonhierarchical,Bolthausen_2009_nonhierarchical,Franz_1992_Ultrametricity,Genovese_2017_Overlap}. More challenging would be to understand the generality or the limitations of corrections  such as (\ref{eq:General:Yk_finite_size},~\ref{eq:General:Delta_2}) in the broader context of mean field models of disordered systems.

\section*{Acknowledgements}
PM would like to thank both Coll\`{e}ge de France and the Institute for Condensed Matter and Complex Systems at the University of Edinburgh for their kind hospitality during this research project.

\bibliographystyle{unsrtpjm}
\bibliography{grem-paper-v5}
\appendix
\section{Derivation of the recursion relations for $G_s(x)$ and $H_s(x)$}
\label{sec:appendix_recursion}	

\subsection{The model in discrete form}

In this appendix we construct the Poisson GREM beginning with a discrete version of the Poisson point process where the energy $E$ is an integer multiple of $\de$, a small energy interval that we will eventually take to zero. As in section~\ref{sec:definition-of-the-poisson-grem} the tree is constructed recursively starting with a single vertex at level $\tau$. We consider an energy scale divided into intervals of size $\de$ with each interval labelled by an integer $i_\tau$ taking values from $-\infty$ to $+\infty$. With each interval we associate a random variable $X_{i_\tau}$ which takes the values zero or one. If $X_{i_\tau}=1$ we
construct an edge from the root vertex and associate an energy, $E=i_\tau \de$, with the edge. These edges we will refer to as level $\tau$ edges. Each level $\tau$ edge connects the level $\tau$ vertex to a level $\tau-1$ vertex. We repeat this construction for each level $\tau-1$ vertex and repeat it again until we reach a vertex at level zero. A vertex at level zero is a leaf node that can be identified with a configuration of the system. A leaf node (or configuration) exists if 

\begin{equation}
	\label{eq:Derivation:Property:XX_X=1}
	X_{i_\tau}X_{i_\tau i_{\tau-1}}\cdots  X_{i_\tau\cdots  i_1}=1
\end{equation}
for the set of integers $\{i_\tau, i_{\tau-1}, \ldots ,i_1\}$ and its energy is

\begin{equation}
	\left(i_\tau+i_{\tau-1}+ \cdots  +i_1\right) \de
\end{equation}
The partition function for this "discrete" Poisson GREM is then

\begin{multline}
	Z_\tau=
		\sum _{i_\tau=-\infty }^{+\infty } \sum _{i_{\tau-1}=-\infty }^{+\infty } 
			\cdots  \sum _{i_1=-\infty }^{+\infty }
		 X_{i_\tau}X_{i_\tau i_{\tau-1}}\cdots X_{i_\tau \cdots  i_1}
		\\ \times \exp \left(-\beta \de \left(i_\tau+i_{\tau-1}+ \cdots +i_1\right) \right)
\end{multline}
It is clear that $ Z_\tau $ satisfies a recursion  

\begin{equation}
\label{eq:Derivation:Z_recursion}
Z_\tau = \sum_{i_\tau = -\infty}^{\infty} X_{i_\tau} e^{-\beta i_\tau \de} Z_{\tau-1}(i_\tau)
\end{equation}
where $ Z_{\tau-1}(i_\tau) $ is the partition function for a sub-tree of height $ \tau-1 $. The sub-tree's single top vertex corresponds to vertex $ i_\tau $ of the main tree. The recursion has initial condition $ Z_0 = 1 $.

The $X_{i_\tau \cdots i_s}$ are the source of quenched disorder in the model. They are all independent random variables (independent for each interval at a given level and between levels), such that

\begin{equation}
	X_{i_\tau \cdots i_s}=
	\begin{cases}		
			1 & \text{with probability } \de \; \rho_s \!\left(i_s \de\right) \\
			0 & \text{with probability } 1-\de \; \rho_s \!\left(i_s \de\right) \\
	\end{cases}
\end{equation}
where $ \tau \ge s \ge 1 $. We will represent the average over all $X_{i_\tau \cdots i_s}$ random variables in the tree by angle brackets $\langle  \ldots \rangle$. 

\subsection{Recursion for the generating function $G_\tau(x)$}

To handle the probability distribution of the partition function it is convenient to introduce a generating function that can be used to generate moments of the partition function,

\begin{equation*}
	G_\tau(x)= \left\langle \exp \left\{-e^{-\beta x}Z_\tau(\beta )\right\}\right\rangle.
\end{equation*}
We can use the recursion~\eqref{eq:Derivation:Z_recursion} to write

\begin{equation*}
G_\tau(x)=\left\langle 
	\prod _{i_\tau=-\infty }^{+\infty } 
	\left[X_{i_\tau} \exp 
	\left\{e^{-\beta(x+ i_\tau \de)} Z_{\tau-1}(i_\tau)\right\} 
		+ (1-X_{i_\tau})
	\right] 
\right\rangle
\end{equation*}
Taking the disorder average then gives the recursion

\begin{equation*}
	G_\tau(x)=\prod _{i_\tau=-\infty }^{+\infty } \left[1+\de \, \rho_\tau( i_\tau \de ) \left\{G_{\tau-1}\left(x+i_\tau \de \right)-1\right\}\right]
\end{equation*}
Taking the limit $\de\to 0$ in such a way that $ i_\tau \de \to E $, a continuous variable, the sums become integrals and we obtain the required recursion 

\begin{equation}
	\label{eq:Derivation:Recursion:Gtau=exp_integral_rho_Gtau-1}
	G_\tau(x)=\exp \left\{\int_{-\infty }^{\infty } \rho_\tau(E) \left[G_{\tau-1}(x+E)-1\right] \dd E\right\}.
\end{equation}
with the initial condition 

\begin{equation}
	\label{eq:Derivation:InitialCondition:G0=exp_exp_beta_x}
	G_0(x)=\exp \left\{-e^{-\beta x}\right\}.
\end{equation}

\subsection{Overlap probability and the generating function $H_\tau^{(p)}(x)$}

We are interested in the probability that $k$ copies of the same tree, each in thermodynamic equilibrium, will be in configurations that have the same edges at level $\tau,$ level $\tau-1$, ... ,and level $p$. This means that they have the first $\tau-p+1$ edges in common starting from the top of the tree. The disorder average of this probability (which we will refer to as the overlap probability) is given by

\begin{multline}
	\label{eq:Derivation:Definition:Qtau=average_Z_to_k}
	Q_\tau^{(k)}(\tau-p+1)=
	\\ \left\langle \frac{\sum_{i_\tau=-\infty }^{+\infty} \cdots \sum_{i_p=-\infty}^{+\infty } 
		X_{i_\tau}\cdots X_{i_\tau \cdots i_p} e^{\left(-k \beta \de\left(i_\tau + \cdots +i_p\right)\right)}
		\left[Z_{p-1}\left(i_\tau,\ldots ,i_p\right)\right]^k}
	{\left[Z_\tau\right]^k}\right\rangle 
\end{multline}
where $Z_{p-1}\left(i_\tau,\ldots ,i_p\right)$ is the partition function for a sub-tree of height $ p-1 $ where the sub-tree's single top vertex corresponds to vertex $\left(i_\tau,\ldots,i_p\right)$ of the main tree. It is given by

\begin{multline*}
	Z_{p-1}\left(\beta ,i_\tau,\ldots ,i_p\right)=\sum_{i_{p-1}=-\infty }^{+\infty } \cdots  \sum _{i_1=-\infty }^{+\infty } X_{i_\tau\text{$\cdots $i}_{p-1}}\cdots
	X_{i_\tau\cdots  i_1}
	\\ \times \exp \left(-\beta  \de\left(i_{p-1}+ \cdots  +i_1\right) \right).
\end{multline*}
Using the integral identity 

\begin{equation}
	\frac{1}{W^k}=\frac{\beta }{\Gamma (k)}\int _{-\infty }^{\infty }e^{-\beta  k x} e^{-e^{-\beta  x}W}\dd x
\end{equation}
one can move the partition function out of the denominator in equation~\eqref{eq:Derivation:Definition:Qtau=average_Z_to_k}. We can then write

\begin{equation*}
	Q_\tau^{(k)}(\tau-p+1)=
	\frac{\beta}{\Gamma (k)}\int_{-\infty }^{\infty } H_\tau^{(p)}(x)  \dd x,
\end{equation*}
where $1\leq p\leq \tau$ and we have introduced a new generating function 

\begin{multline}
	\label{eq:Derivation:Definition:Htau=average_sum_phi_derivative}
	H_\tau^{(p)}(x)=
	\Bigg\langle \sum_{i_\tau=-\infty }^{+\infty } \cdots \sum_{i_p=-\infty }^{+\infty } 
	X_{i_\tau}\cdots X_{i_\tau \cdots i_p} 
	e^{-k \beta \de\left(i_\tau + \cdots +i_p\right)}
	\\ \times 
	\left[e^{-\beta x} Z_{p-1}\left(i_\tau\ldots i_p\right)\right]^k
	\exp \left(-e^{-\beta x} Z_\tau\right)  \Bigg\rangle 
\end{multline}
One can express the generating function $H_t^{(p)}(x)$ in recursive form. Averaging on the $X_{i_\tau}$ this gives, using \eqref{eq:Derivation:Z_recursion} and after some simplification, the recursive form

\begin{multline*}
	H_\tau^{(p)}(x)=
	\sum _{i=-\infty }^{+\infty }  \de \, \rho _\tau\!\left(i \de\right) 
		H_{\tau-1}^{(p)}\left(x+i \de\right)
	\\ \times
	\prod_{
		\substack{j=-\infty  \\
			\left(\neq i \right) \\}
		}^{\infty } 
	\left[1+\de \, \rho_\tau (j \de) \left\{G_{\tau-1}\left(x+j \de\right)-1\right\}\right].
\end{multline*}
Lastly, taking the $\de\to 0$ limit in the appropriate way we obtain a recursive integral equation.

\begin{equation}
	H_\tau^{(p)}(x)= G_\tau(x)\int _{-\infty }^{\infty }\rho_\tau(E) \; H_{\tau-1}^{(p)}(x+E) \dd E.
\end{equation}
The initial condition is obtained by taking $\tau=p$ in equation~\eqref{eq:Derivation:Definition:Htau=average_sum_phi_derivative}. After taking the disorder average this becomes

\begin{multline}
H_p^{(p)}(x)=  \, e^{-k \beta x} \sum _{i=-\infty }^{+\infty }  \de \, \rho_p (i \de) 
\left\{\left[\frac{e^{\beta  x }}{\beta }\frac{\dd}{\dd x}\right]^k G_{p-1}\left(x+i \de\right)\right\}\\
\times \prod_{
	\substack{j=-\infty  \\
		\left(\neq i \right) \\}
}^{\infty } 
\left[1+\de \, \rho_p(j \de) \left\{G_{p-1}\left(x+j \de\right)-1\right\}\right].
\end{multline}
Taking the $\de\to 0$ limit in the appropriate way we obtain

\begin{equation}
	H_p^{(p)}(x)= G_p(x)e^{-k \beta  x }\int _{-\infty }^{\infty }\rho_p(E) \left[\frac{e^{\beta  x }}{\beta }\frac{\dd}{\dd x}\right]^k G_{p-1}(x+E) \dd E.
\end{equation}

\section{Relationship between the standard GREM and the Poisson GREM}
\label{sec:standard-vs-poisson-grem}
In this appendix we show how to connect the Poisson GREM that we
consider in the present paper with the GREM as it was defined earlier in the literature \cite{Derrida_1985_generalization,Derrida_1986_Solution,Ruelle_1987_Mathematical,Cook_1991_Finite,Bovier_2004_Derridas,Bovier_2004_Derridasa}. In contrast to the Poisson GREM,
the standard GREM is a regular tree in the sense that each vertex at a
given level $s$ is  s connected to the same number $\alpha_s^N$ of
vertices at the lower level $s-1$. Moreover associated to  each edge $b$
connecting a  vertex at level $s$ to a vertex at level  $s-1$, there is
an energy $E_b$ distributed according to a gaussian distribution
$$\widetilde{\rho}(E_b) = {1\over \sqrt{\pi N a_s}} \exp \left[-{E_b^2
	\over N a_s} \right] $$
For a GREM of height  $\tau$, the  model is fully  specified  by the $2
\tau $ parameters $\alpha_s$ and $a_s$ for $1 \le s \le \tau$.

To relate the standard GREM to the Poisson GREM, one simply needs to
remember that (for the choices of the parameters $\alpha_s$ and $a_s$
which give rise to  either gradual or simulatenaous freezing) the only
part of the distribution $\widetilde{\rho}(E_b) $ which matters in the
large $N$ limit is
the part located near the energy
$$E_s^* = N \sqrt{a_s \log \alpha_s}$$
Looking at   the energies $E$ of the edges connecting all the vertices at
level $s-1$ to a given vertex at level $s$, those   which lie close to
$E_s^*$ are distributed according to a Poisson process
of intensity
 \[\alpha_s^N \widetilde{\rho}(E) \simeq C_s \exp \left[ A_s (E-E_s^*) -
{B_s \over N} (E-E_s^*)^2 \right]\]
with
\[A_s = 2 \sqrt{\log \alpha_s \over a_s} \ \ \ \ \  ;  \ \ \ \ \   B_s =
{1 \over a_s} \ \ \ \ ; \ \ \ \ C_s = {1 \over \sqrt{ N \pi a_s} } \]
So up to a global shift of energy by $E_s^*$ this is precisely the Poisson
GREM as defined in section 2.

As discussed in our earlier paper on the REM \cite{Derrida_2015_Finite}, we expect the leading finite size corrections in free energy between the Poisson GREM and the standard GREM to coincide.
\end{document}